\documentstyle[twocolumn,aps,floats]{revtex}

\begin{document}

\draft \tolerance = 10000

\setcounter{topnumber}{1}
\renewcommand{\topfraction}{0.9}
\renewcommand{\textfraction}{0.1}
\renewcommand{\floatpagefraction}{0.9}
\newcommand{\br}{{\bf r}}

\twocolumn[\hsize\textwidth\columnwidth\hsize\csname
@twocolumnfalse\endcsname

\title{Will the  Population of Humanity in the Future be Stabilized? }
\author{L.Ya.Kobelev, L.L.Nugaeva \\
Department of  Physics, Urals State University \\ Lenina Ave., 51,
Ekaterinburg 620083, Russia  \\ E-mail: leonid.kobelev@usu.ru}

\maketitle

\begin{abstract}
 A phenomenological theory of growth of the population of humankind is
proposed. The theory based on the assumption about a multifractal nature
of the set of number of people in temporal axis and contains control
parameters.  In particular cases the theory coincide with known Kapitza ,
Foerster, Hoerner phenomenological theories .
\end{abstract}

\pacs{ 01.30.Tt, 05.45, 64.60.A; 00.89.98.02.90.+p.} \vspace{1cm}

]

\section {Introduction}
The problem of mankind population growth is the one of the global problems
concerning the development of the mankind and its future. Will the
demographic explosion existing now at the mankind population of the whole
world stop as already it has stopped in the most developed countries? Will
the population of the Earth will be stabilized as it follows from
S.Kapitza's theory (see \cite{1})near 12 billions, or it will continue its
growth at slower rate? What are  the driving parameters that govern the
development of mankind (such as presents in non-linear open system
(see\cite{2})? Are these parameters genetically predetermined or can they
be changed and controlled by means of human activity? Stabilization of the
mankind population of earth in the future is a sad prospect for mankind,
because the absence of the numerical increasing of any biological
population almost always leads, early or late, to cessation of any
development (the examples are many species of insects, e.g. termites,
frozen in development for millions years). Hence, the appearance of more
active biological species becomes in that case quite probable. These
active form of life will dominate the mankind and may force it out from
its present ecological niche. The aim of the present paper is to introduce
a new parameter in the phenomenological theories of humankind growth
describing the development of mankind population if regard humankind as a
large non-linear multifractal system. Namely, the fractal dimension of the
whole mankind number in arbitrary moment of time - the local fractional
dimension ${d(t)}$ (the fractional dimension for whole population of
mankind). The introducing  of this parameter allows  to receive, as a
special cases, the results of theories (see \cite{1}),\cite{3}-\cite{4}),
and several new scenarios of the mankind's future as well. Alongside with
probable increasing of the mankind population, the scenario of ruin of the
present civilization - diminution of its number down to zero - is shown to
be possible. We note, that correct analysis of dynamics of multifractal
sets (see\cite{5}) requires the introduction of the mathematical concept
of fractional derivatives (see \cite{6}-\cite{7}), which allow to take
into account partly the memory of system about the past (in this case it
is the memory, that includes genetic memory of mankind about its past
development).

\section{Fractional Dimension Quantity of Mankind on Time axis}

We assume that the dynamics of the human population can be described
within the framework of fractal geometry concepts and mathematical
formalism of fractional differential equations (see \cite{6}-\cite{7}).
For this purpose let us consider the set of all people as a multifractal
set $N(t)$ $(0\leq t\leq \infty)$ consisting of $N(t)$ of elements at the
given time $t$. Following the Kapitza theory (see \cite{1}), assume for a
certain interval of time the existence of a self-similarity of this set
characterized by its fractal dimension and introduce a local fractional
dimension (LFD) $d(t)$ which describes the fractal properties of the set
at the time $t$. This local fractional dimension is determined by those
variables and their functions (these variables will be defined later) ,
that are treated as the driving (control) parameters of the human
community development. Among such parameters there can be parameters of
genetic origin  (for example, probably, density of the population in
cities per unit urban area, etc.) and "external" parameters (e.g., a
possibility of supplying the mankind population of earth by necessary
amounts of food, water and energy or quick climate changes or pollution of
the environment and the atmosphere, etc.). We shall characterize  an
alteration of the mankind population $N(t)$  over a small time interval by
the generalized fractional Riemann - Liouville derivative [7] (which
coincides with the usual Riemann - Liouville derivative if $d$=const.)

\begin{equation}\label{eq1}
D_{+,t}^{1+ \nu(t)}N(t)=\frac{{\partial ^\alpha}}{{\partial
t^\alpha}}\int\limits_0^t {\frac{{N(t')dt'}}{{\Gamma(\alpha- d(t'))(t -
t')^{d(t') -\alpha+ 1}}}} ,
\end{equation}
\begin{eqnarray}\label{equ1} \nonumber
d(t)=1+\nu(t)>0
\end{eqnarray}

\begin{equation}\label{eq2}
\nu(t)\equiv\nu(X_{1}(t),X_{2}(t)...X_{i}(t)), i=1,2,..., \alpha=\{d\}+1
\end{equation}

In (\ref{eq1}) $\nu(t)$ is the fractional quantity and defines the
differences between the derivatives of integer order and fractional
derivatives (\ref{eq1}) thus being the driving parameter for the growth of
mankind as a whole, $\{d\}$ is equals to the integer part of $d(t)>0$
$(\alpha-1\leq d(t)<\alpha)$ , $\alpha=0$  for $d<0$ , and the set $X_{i}$
are the control parameters determining all external and internal
influences on the mankind population growth .

The explicit information about the function $\nu$ and, hence, about LFD
can be obtained only after a careful investigations and processing the
statistical data of different events, circumstances and situations impact
on the development of a human population. The fractional derivative using,
as defined in (\ref{eq1}), instead of the integer first derivative allows
to introduce and take into account an obvious thing - a certain kind of
mankind's memory of the past and memory about the development rates over
the past years (integration with a weight function over all times till $t$
beginning from a fixed moment). It gives a way of considering of different
parameters $X_{i}$ that influence at the mankind's development by means of
LFD's dependence upon them. The present theory, as well as (\cite{1}), is
a phenomenological theory and the exact definition of function $\nu(t)$
form is beyond its scope. We note that it has sense to consider
$\nu(t)\neq 1$, apparently, only for times greater than $T_{1}'$ because
before a certain time, introduced in (\cite{1}, \cite{3}-\cite{4}) (the
time of demographic transition $T=T_{1}'$), the theories mentioned
describe the empirical data about the number and progress of the mankind
population quite well. Moreover, we shall restrict ourselves in this paper
to analyzing growth of the mankind population for three special cases of
$\nu(t): \nu(t)=0,\nu(t)=1$ and $\nu(t)=-1$.

\section {Foerster,Hoerner, Kapitza Theories}
It is known, that growth of the population of earth from ancient times
untill now is well approximated by an empirical relation suggested by
Foerster Von H. (see \cite{3}) and improved  by H. von Xoerner (see
\cite{4})

\begin{equation}\label{eq3}
N(t)=\frac{{C}}{{T_{1}-T'}},
\end{equation}

\begin{equation}\label{eq4}
\frac{{d(N(t))}}{{dt}}=\frac{{C}}{{(T_{1}-T)^{2}}}
\end{equation}

$C=2^{.}10^{11}, T_{1}=2025$

The reasonable generalization (\ref{eq3}),(\ref{eq4}) for future time
(suitable for $T>T_{1}'$ that not gives as result an infinite value $N(t)$
for values $T=T_{1}$) was given in the phenomenological theory of the
population S.Kapitza (see \cite{1}) with the help of introducing of the
mean people's lifetime   $\tau$ ($\tau$ =42 of years). In this theory the
relation for $N(T)$ (with C'=1,86$^{.}10^{11}, T_{1}'=2007$) takes place

\begin{equation}\label{eq5}
 \frac {{\partial N(t)}}{{dt}}= \frac{{C'}}{{(T_{1}'-T)^{2} + \tau ^{2}}}
\end{equation}

From (\ref{eq5})  the restriction of the mankind population of earth by
quantity $14^{.}10^{9}$  follows. Unfortunately, the theory of the
population S.Kapitza does not take into account neither exterior nor
interior control parameters (even in simple form) basing only  at the
self-similarity of growth of a population of the people.

\section {Generalization of Foerster, Hoerner and  Kapitza Theories for
Multifractal  Set of a Quantity of Mankind $N(t)$}

Let us assume the hypothesis about a fractal nature of set $N(t)$
(maintaining assumption about selfsimilar the sets $N(t)$). In that case
derivative on time  $\frac{\partial}{\partial t}$  in equation (\ref{eq5})
should be substitute for fractional derivative $D^{1+\nu(t)}_{+,t}$. This
operation take into account the memory of mankind about the past
development. The right part (\ref{eq5}) must be changed too for including
in (\ref{eq5}) an influence of the fractal dimension. So, instead of
(\ref{eq5}), obtain an equation
\begin{equation}\label{eq6}
D_{+ ,t}^{1 + \nu(t)}N(t)= \frac{{C'}}{{|T_{1}'-T|^{2+ \nu(t)}+ \frac{{2+
\nu(t)}}{{2}}\tau ^{2 + \nu (t)}}}
\end{equation}

The equation (\ref{eq6}) can be considered as the basic equation of the
phenomenological theory of development mankind's population offered in
this paper. The selection of different functions of fractional corrections
for $\nu(t)$   allows to estimate character of changing $N(t)$  as
functions of time.

\section {Forecasts of the Future Development of Mankind Population for
Special Cases of  the Fractional Dimension Choice}

Some simple special cases forecasting of growth of the mankind's
population of earth are considered below at the basis of the equation
(\ref{eq6}). The meanings a fractal dimension, for simplicity , chosen as
integer.

a. Case $\nu(t)=0$

At $\nu(t)=0$ the fractional derivative $D$ coincides with
$\frac{\partial}{\partial t}$ derivative and the equation (\ref{eq6})
coincides with equation (\ref{eq5}) (hence, (\ref{eq6}) includes S.Kapitza
theory (see \cite{1}) as a special case). It corresponds, probably, a
compensation the positive and the negative control parameters $X_{i}$
(including exterior and "interior" parameters), driving population of
mankind.

b. Case $\nu(t)\rightarrow {-1}$

The selection $\nu(t)\rightarrow{-1}$  corresponds to a dominance of the
negative tendencies in the future development of mankind and it is
stipulated, for example, occurrence of irreversible changes in molecules
DNA (owing to epidemic AIDS, etc.), irreversible cosmic cataclysmic (for
example, drop on earth of meteorites of huge mass), impossibility for
mankind to cope with negative factors of biogeozinos results in by effects
of development of a technical civilization. In this case the equation
(\ref{eq6}) reduce to equation
\begin{equation}\label{eq7}
N(t)=\frac{{2C'}}{{2|T_{1}'-T|+\tau}}
\end{equation}

From (\ref{eq7}) the presence of the maximum of number of mankind follows
at  $T=T'$ (i.e. in 2007) and it is equal $8,86^{.}10^{9}$ (if $\tau$ is
not changed). After transition through a maximum $N(t)$ the number of
mankind decreases (if the scenario will not vary) and in 2107 year $N(t)$
will be equal $1,54^{.}10^{9}$. By the year 3007 the number of mankind
will decrease down to $182^{.}10^{6}$, i.e. the mankind population of all
earth will be equal to number of the people is occupying a dozen of large
modern cities. The complete degenerating of mankind, as a result of
decreasing of mankind population in the considered scenario and subsequent
leaving mankind from biological arena  may be expected through millions
years ( not large time from biological point of view). By then population
the mankind of earth will decrease up to several thousand.

c. Case $\nu(t)=-2.$

The equation (\ref{eq6}) for this case is transformed into an integral
equation at maintenance of the general tendency to accumulation of the
negative factors resulting in to negative value $d(t)$ . At negative
values $d(t)$ the integral equation gives the prompt diminution of mankind
population and extinction of existence of mankind as a species is follows.
So, at $d(t)\rightarrow{-1}$ equations (\ref{eq6}) transforms to a
\begin{equation}\label{eq8}
\int\limits_{T'}^T {N(t)dt}  = C'
\end{equation}

supposing the absence  the mankind at the earth (in that case equation
(\ref{eq8}) has no solution for $N(t)\neq{0}$  thou for $\nu>-2$ solution
of equation (\ref{eq6}) is exist (so $N(t)\rightarrow{0}$  for
$\nu\rightarrow{-2}$). The time interval (necessary for disappearance
mankind) is determined in this case by time for which $d(t)$ will transfer
from value equal to unity (status of mankind now) to value equal to
negative unity. This time interval can be very short: from several years
(cosmic cataclysmic) to about several centuries (virus pandemia with
lethal change of a heredity, increasing of the mean temperature of earth
on some degrees owing to throw out of carbonic gas etc.). It is necessary
to pay attention on possibility of change of the negative scenario of
mankind development  stipulated by appearance and dominance of the
positive control factors $X_{i}$ (including those factors that due to
conscious activity of mankind).  In that case the inevitability of
diminution of mankind population and destruction of mankind are not
inevitable.

d. case $\nu=1.$

We shall consider the optimistic scenario  of change of mankind
population. It relevant to a dominance of positive driving parameters:
$d(t)>1.$ So, e.g. for $d(t)=2$ $(\nu =1)$ from (\ref{eq6}) the equation
follows
\begin{equation}\label{eq9}
\frac{{\partial^{2}}}{{\partial t^{2}}}N(t)=\frac{{C'}}{{|{T'- T}|^{3} +
1,5\tau^{3}}}
\end{equation}

The precise solution of an initial value problem of this equation is
unwieldy, so we shall note , that at $T>>\tau$ the quantity $N(t)$  will
increase faster then first degree of time ($N(t)\sim(T-T_{1})$). The
mankind population is increase  and it characterizes in this case by
following form (if there will not be includes appearance of a factors of
conscious mankind activity which change the scenario and restricts
unlimited growth of population)

\begin{equation}\label{eq10}
N(t)|^{t \to \infty}\sim\frac{C'(T-T_{1}')}{2,29\tau^{2}} \ln \frac{(T -
T_{1}')}{\tau}
\end{equation}

So, at 3000 years, at the rate of increasing of the population  defined by
(\ref{eq10}) (with the allowances that the corrections to $N(t)$ of value
$ln [\frac{{ (T-T_{1}')}}{{ \tau}}]$ are dropped) population of mankind
will increase up to ~150 billions. That is improbable large, though, but
it is may be not unreasonable because of future technical possibilities of
mankind (probably, this number  is the upper number for existence mankind
population occupying the earth). For fractional values LFD $d(t)$,
increasing or decreasing of a population of the mankind will be
characterized by intermediate dependencies between the received for the
whole values $d(t)$. In case of a periodic dependencies $d(t)$ from time
the population of the world will change periodically depending on a
concrete choice of $d$ and $N(t)$ and will not be a monotonous function of
time.

The examples are considered allow to determine the interval of change of
fractional dimension $d(t)$ in reason boundary for number of quantity
mankind in future: $-1<d<2$. The boundary values $(d=-1, d=2)$ are result
in or to ruin of mankind, or to  so large mankind population that Earth
may not endure. The last case  must result in to change of scenario and it
consist in the change in correlation between positive and negative
components of control parameters $X_{i}$ towards increasing of influence
of negative parameters.

 Let the basic equation (\ref{eq6}) is replaced by generalization of the
S.Kapitza equation (\ref{eq5}). Basic equation in that case reads

\begin{equation}\label{eq11}
\frac{{\partial^{1 + \nu(t)}}}{{\partial t^{1+\nu(t)}}}n(t)= K\sin^{2}
\frac{{n(t)}}{{K}}=\frac{{1}}{{K}},
\end{equation}
\begin{eqnarray}\label{equ2} \nonumber
n=\frac{{N(t)}}{{K}}, t=\frac{{T -T_{1}'}}{{\tau}}, K=\sqrt{{C'\tau^{-
1}}}
\end{eqnarray}

It gives for $N(t)$ the qualitative effects analogous to results obtained
from the equation (\ref{eq6}). So, in this connection, a selection for
describing the future increasing the population of mankind by equation
(\ref{eq6}), as the basis, or equation (\ref{eq11}), containing, as well
as the equation (\ref{eq6}), driving parameters $X_{i}$ in fractional
dimension $d(t)$, is not simple. As one of advantages of use of the
equations (\ref{eq6})  or (\ref{eq11}) for describing demographic problems
(with some of them the mankind already has confronted now) we shall stress
an opportunity of insert and account in the theory many factors (such as ,
incurable illnesses, natural cataclysmic etc.) defining future of mankind
as a result of influences the control parameters $X_{i}$ (included in the
dimension $d(t)$ of fractal set for number's distribution of  the people
in the time axis).

\section {Conclusions}

1.The main purpose of this paper was to analyze possibility of introducing
mankind's population driving parameters $X_{i}$ in the phenomenological
theories of the mankind population of the earth (\cite{1}, \cite{3},
\cite{4}) by method of attributing to set of the people $N(t)$ the
fractional dimension $d(t)$, depending from these parameters. At a choice
$d(t)=1$ the numbered theories are a special case of this phenomenological
theory. The consideration of examples with integer meanings of $d(t)$ is
caused only by their mathematical simplicity and gives a reasonable
meaning of fractal dimensions ($-1<d<2$) for describing the time
dependence of population of mankind.

2. In case when the interpretation the fractal properties of set of the
homosapience given in this paper corresponds to a reality, ( more real
cases correspond to fractional meanings  $d(t)$)  the future of mankind is
not so mournful as in the case of the S.Kapitza theory (see \cite{1}) and
the exposition of change of number of mankind within the framework of the
phenomenological theories of the population can be reduced to a selection
of control  parameters and filling them by the concrete contents.

3. The change of number of mankind (described in the framework a
phenomenological theories of the population) can be adjusted by such
choice of control parameters (and filling the fractal dimension $d(t)$ by
the concrete contents  of dependencies of them)  at which the population
of mankind will grow so slowly, that overpopulation and the problems
connected with it will do not arise in the foreseen future. Last will
allow the theory be more realistic for predicting and menaging the future
growth population of mankind as one of biological species occupying our
world.

4. The growth of mankind population regulation (included in the parameters
$X_{i}$)  will allow to avoid degenerating of mankind and to keep as much
as long  time the main ecological niche at Earth occupied by mankind. Last
will give time for more realistic forecasting of the future mankind as one
of biological kinds occupying our world.


\begin{references}

\bibitem{1}Kapitza S.P., Uspehi Fizicheskih Nauk (Russia), 1966, vol.166, No.1,
pp.63-79; Kapitza S.P., \emph{How Many People Lived, Live and are to Live
in the World. An essay on the theory of growth of humankind}, Moscow,
Inst. Phys. Problem RAS, 1999,238p.
\bibitem{2} Klimontovith Yu.L., \emph{Statistical Theory of Open Systems. Vol.1},
Moscow, Yanus, 1995, 686p. (in Russian); Kluwer Academic Publishers,
Dordrecht, 1995; Klimontovich Yu.L., \emph{Statistical Physics of Open
systems. Vol.2}, Moscow, Yanus, 1999, 450p. (in Rusian).
\bibitem{3} Foerster, Von H. et al., Science, v.132, p.1291 (1960)
\bibitem{4} Hoerner, von SJ., British Interplanetary Society, v.28, 691, (1975)
\bibitem{5} Mandelbrot B., \emph{Fractal Geometry of Nature}, W.H.Freeman, San Francisco, 1982
\bibitem{6} Samko S.G, Kilbas A.A., Marithev I.I., \emph{Integrals and Derivatives
of the Fractional Order and Their Applications}, (Gordon and Breach, New
York, 1993).
\bibitem{7}Kobelev L.Ya, \emph{Fractal Theory of Time and Space}, Ekaterinburg,
Konros, 1999, 136p. (in Russian);Kobelev L.Ya., What Dimensions Do the
Time and Space Have: Integer or Fractional? arXiv:physics/0001035; Kobelev
L.Ya., Can a Particle's Velocity Exceeds the Speed of Light in the Empty
Space? arXiv:gr-qc/0001042; Kobelev L.Ya.,Physical Consequences of Moving
Faster than Light in Empty Space, arXiv:gr-qc /0001043 ;Kobelev L.Ya.,
Multifractality of Time and Space, Covariant Derivatives and Gauge
Invariance,arXiv:hep-th/ 0002005; Kobelev L.Ya.,Generalized Riemann
-Liouville Fractional Derivatives for Multifractal Sets, arXiv:math.
CA/0002008,; Kobelev L.Ya., The Multifractal Time and Irreversibility in
Dynamic Systems, arXiv:physics/0002002; Kobelev L.Ya., Is it Possible to
Transfer an Information with the Velocities Exceeding Speed of Light in
Empty Space?,arXiv: physics/ 0002003; Kobelev L.Ya., Maxwell Equation,
Shroedinger Equation, Dirac Equation, Einstein Equation Defined on the
Multi fractal  Sets of the Time and the Space, arXiv:gr-qc/0002003

\end{references}
\end{document}